\documentclass[a4paper,fleqn,usenatbib]{mnras}
\usepackage{amsmath}
\usepackage{epsfig} 
\usepackage{graphicx}	
\usepackage{amssymb}	

\newcommand{\be}{\begin{equation}}
\newcommand{\e}{\end{equation}}
\newcommand{\bear}{\begin{eqnarray}}
\newcommand{\ear}{\end{eqnarray}}

\newcommand{\hmpc}{{\, h^{-1}\, {\rm Mpc}}}

\def\aj{AJ}
\def\apj{ApJ}
\def\apjs{ApJS}

\def\mnras{MNRAS}
\def\aap{A\&A}

\def\apjs{ApJS}
\def\apjl{ApJ Letters}

\title[Galaxy morphology and large-scale environment] {How much a
  galaxy knows about its large-scale environment?: An information
  theoretic perspective}
    
\author[Pandey, B. and Sarkar, S.] {Biswajit Pandey\thanks{E-mail:
    biswap@visva-bharati.ac.in} and Suman
  Sarkar\thanks{E-mail:sumansarkar.rs@visva-bharati.ac.in}
  \\ Department of Physics, Visva-Bharati University, Santiniketan,
  Birbhum, 731235, India\\ }
   
 \date{\today}

 \pubyear{2016}
  
\begin{document}
\label{firstpage}
\pagerange{\pageref{firstpage}--\pageref{lastpage}}      
\maketitle
       
\begin{abstract}

The small-scale environment characterized by the local density is
known to play a crucial role in deciding the galaxy properties but the
role of large-scale environment on galaxy formation and evolution
still remain a less clear issue. We propose an information theoretic
framework to investigate the influence of large-scale environment on
galaxy properties and apply it to the data from the Galaxy Zoo project
which provides the visual morphological classifications of $\sim 1$
million galaxies from the Sloan Digital Sky Survey. We find a non-zero
mutual information between morphology and environment which decreases
with increasing length scales but persists throughout the entire
length scales probed. We estimate the conditional mutual information
and the interaction information between morphology and environment by
conditioning the environment on different length scales and find a
synergic interaction between them which operates upto at least a
length scales of $ \sim 30 \hmpc$. Our analysis indicates that these
interactions largely arise due to the mutual information shared
between the environments on different length scales.

\end{abstract}


\section{Introduction}
Understanding the formation and evolution of galaxies is one of the
most challenging problems in cosmology. It is now quite well known
that the galaxy properties depend on the environment. The elliptical
galaxies are known to preferentially reside in rich clusters whereas
the spiral galaxies are mostly distributed in the fields
\citep{hubble, zwicky, dress}. Analysis of the two-point correlation
function of galaxies \citep{will, brown, zehavi} also suggests that
the ellipticals are strongly clustered as compared to the spirals. The
genus statistics of the red galaxies which are predominantly
ellipticals show a shift towards a meatball topology \citep{hoyle1,
  park1} indicating their preference for the high density
environments. A study of the filamentarity of the galaxy distribution
\citep{pandey2} in the SDSS DR1 indicates that the ellipticals
preferentially inhabit the nodes where the filaments intersect whereas
the spirals are sparsely distributed along the filaments. Besides
morphology, many other galaxy properties such as luminosity, colour,
star formation rate, star formation history, stellar mass, size,
metallicity and AGN activity are known to strongly depend on the
environment \citep{davis2, guzo, zevi, goto, hog1, blan1, einas2,
  kauffmann, mocine, bamford, koyama}.

In the current paradigm, the dark matter first collapses
hierarchically into halos and the baryons settle down later at the
centres of these halos to form galaxies by cooling and condensation
\citep{white1978}. In the halo model \citep{neyman, mo, ma, seljak,
  scocci2, cooray, berlind, yang1}, all galaxies are believed to form
and reside in virialized dark matter halos. The halo model postulates
that the number and type of galaxies residing in a dark matter halo
are entirely determined by its mass. If these halos evolve in
isolation, the galaxy properties are largely determined at birth by
the initial conditions at the locations where they formed. But in the
hierarchical model, smaller halos merge to form bigger halos and the
galaxy properties evolve according to the nature of their mergers. The
environmental effects such as ram pressure striping and different
types of galaxy-galaxy interactions are thus expected to play a
crucial role in the formation and evolution of galaxies.

Traditionally, the environment of a galaxy is characterized by the
local density which describes the neighbourhood of the host halo where
the galaxy is embedded. However the host halos themselves may be
embedded in filaments, sheets or clusters and many of the properties
of the host halos such as their masses, shapes and spins are
determined by their large-scale cosmic environment \citep{hahn1}. So
in principle the large-scale environment can indirectly influence the
properties of the galaxies. Using SDSS DR7 data, \citet{lupa} find a
significant dependence of the properties of late type brightest group
galaxies on their large-scale environment.  A study of the star
formation rates in compact groups from the SDSS DR7 \citep {scudder}
find significant difference in the star formation rates between
isolated groups and the groups which are embedded in superstructures.
\citet{pandey3} analyzed the filamentarity of the galaxy distribution
from the SDSS DR5 and find that the average filamentarity of the star
forming galaxies are higher than the red galaxies. \citet{darvish}
find that the presence of filamentary environment elevates the
fraction of star forming galaxies in the past at redshift $z \sim
1$. \citet{filho} find that $\sim 75\%$ of the extremely metal poor
galaxies reside in sheets and voids. \citet{parkchoi} analyzed the
galaxies from the SDSS DR4 to find that the role of the large-scale
density in determining galaxy properties is minimal once luminosity
and morphology are fixed.  \citet{yan} find no dependence of the
galaxy properties on the tidal environment of large scale structures
in the SDSS DR7. At present, the roles of the large-scale environment
in determining the galaxy properties are less clear and there is no
consensus on this issue.

In this Letter we propose an information theoretic framework to study
the dependence of the galaxy properties on the large-scale
environment. We apply this method to the Galaxy Zoo project
\citep{lintott1,lintott2} to investigate if there is any correlation
between the galaxy morphology and the large-scale environment. More
than $200,000$ volunteers in the internet participated in this project
and classified $\sim 1$ million galaxies according to their morphology
by visual inspection. The morphological classification by direct
visual inspection avoids many of the potential biases associated with
the proxies for morphology. Thus the Galaxy Zoo provides an unique
opportunity to test the large scale environmental dependence of the
morphology of the galaxies. The method proposed here can be also
applied to test the large scale environmental dependence of any other
galaxy properties.

A brief outline of the Letter follows. In section 2 we describe the
method of analysis followed by a description of the data in section
3. We present the results and conclusions in section 4.

\section{METHOD OF ANALYSIS}

We construct a galaxy sample in a cubic region containing $N$ galaxies
with known morphology. We divide the cubic region into $d \, h^{-1}\,
{\rm Mpc} \times d \, h^{-1}\, {\rm Mpc} \times d \, h^{-1}\, {\rm
  Mpc}$ three dimensional rectangular grids. The entire sample is now
divided into a number of regular cubic voxels of grid size $d \,
h^{-1}\, {\rm Mpc}$. Let $N_{d}$ be the number of resulting voxels for
a grid size of $d$. Our sample consists only the spiral and the
elliptical galaxies and does not include the galaxies with uncertain
morphology. One can count the number of different types of galaxies in
each of the $N_{d}$ voxels for grid size $d$. Let $(n_{s})_{i}$ and
$(n_{e})_{i}$ respectively be the number of spiral and elliptical
galaxies residing in the $i^{th}$ voxel then
$n_{i}=(n_{s})_{i}+(n_{e})_{i}$ is the total number of galaxies in the
$i^{th}$ voxel. Summing over all the $N_{d}$ voxels gives
$\sum^{N_{d}}_{i=1} (n_{s})_{i}=N_{s}$, $\sum^{N_{d}}_{i=1}
(n_{e})_{i}=N_{e}$ and $\sum^{N_{d}}_{i=1} n_{i}=N$ where $N_s$, $N_e$
and $N$ are the total number of spirals, total number of ellipticals
and the total number of galaxies in the sample respectively.  We
define two discrete random variables $X$ and $Y$ with probability
distributions $P(X)$ and $P(Y)$ respectively.  $P(X_i)=\frac{n_i}{N}$
is the probability that a randomly drawn galaxy resides in the
$i^{th}$ voxel. $P(X)$ has a total $N_{d}$ outcomes. $P(Y_j)$ is the
probability that a randomly chosen galaxy is spiral or elliptical and
it has $2$ outcomes given by $P(Y_1)=\frac{N_{s}}{N}$ for spiral and
$P(Y_2)=\frac{N_{e}}{N}$ for elliptical. We vary the grid size $d$
within a suitable range and estimate the probability distributions
$P(X)$ in each case.  It may be noted here that we consider a cubic
region for our analysis as it allows us to use the entire sample while
dividing it into different number of rectangular voxels for different
grid sizes and keep the total number of galaxies same in each case.

\subsection{The mutual information between morphology and environment}
In information theory, the information entropy $H(x)$
\citep{shannon48} is the average amount of information required to
describe a random variable $x$ and is defined as,
\begin{eqnarray}
H(x) & = & - \sum^{m}_{i=1} \, p(x_i) \, \log \, p(x_i)
\label{eq:shannon}
\end{eqnarray}
where $p(x_i)$ is the probability of $i^{th}$ outcome and $m$ is the
total number of outcome. Here the base of the logarithm is arbitrary
and we chose it to be 10 for the present work.

For the pair of random variables $(X,Y)$ with joint probability
distribution $P(X,Y)$ the joint entropy $H(X,Y)$ is defined as,
\begin{eqnarray}
H(X,Y) & = & - \sum^{N_{d}}_{i=1} \sum^{2}_{j=1} \, P(X_i,Y_j) \, \log\, P(X_i,Y_j)
\label{eq:joint1}
\end{eqnarray}
Here $X$ and $Y$ are the random variables introduced earlier. The
joint probability that a randomly selected galaxy resides in the
$i^{th}$ voxel and is spiral or elliptical is given by $P(X,Y) =
P(Y|X) P(X)$ where $P(Y|X)$ is the conditional probability that the
randomly selected galaxy is spiral or elliptical given that it resides
in the $i^{th}$ voxel. This gives $P(X_i,Y_j)=\frac{(n_{s})_{i}}{N}$
for $j=1$ (spiral) and $P(X_i,Y_j)=\frac{(n_{e})_{i}}{N}$ for $j=2$
(elliptical).

The mutual information $I(X;Y)$ between $X$ and $Y$ is then defined as,
\begin{eqnarray}
I(X;Y) & = & \sum^{N_{d}}_{i=1} \sum^{2}_{j=1} \, P(X_i,Y_j) \, \log\, \frac{P(X_i,Y_j)}{P(X_i)P(Y_j)} \\
\nonumber & = & H(X)+H(Y)-H(X,Y)
\label{eq:mutual1}
\end{eqnarray}

where $H(X)$ and $H(Y)$ are the individual entropies associated with
the variable X and Y respectively. $I(X;Y)\geq 0$ as $H(X,Y) \leq
H(X)+H(Y)$ with equality only if the random variables $X$ and $Y$ are
independent. The mutual information is also symmetric in $X$ and $Y$
i.e.  $H(X,Y)=H(Y,X)$. It should be noted that $I(X;X)=H(X)$ provides
the self information. We vary the grid size and estimate $I(X;Y)$ in
each case.

The mutual information measures how much one random variable tells us
about another. It can be thought of as the reduction in uncertainty
about one random variable given the knowledge of another. A high value
of mutual information corresponds to large reduction in uncertainty
and low value indicates a small reduction in it. A zero mutual
information indicates that the two random variables are independent.

\subsection{The mutual information between environments on different scales}
We also estimate the mutual information between the environments on
different length scales. We define another random variable $\bar{X}$
analogous to $X$ which characterizes the environment of the galaxy on
an another scale. The mutual information between environments on
different length scales $I(X;\bar{X})$ is similarly defined as,
\begin{eqnarray}
\nonumber I(X;\bar{X}) & = & \sum^{N_{d_1}}_{i=1} \sum^{N_{d_2}}_{k=1} \, P(X_i,\bar{X}_k) \, \log\, \frac{P(X_i,\bar{X}_k)}{P(X_i)P(\bar{X}_k)} \\ & = & H(X)+H(\bar{X})-H(X,\bar{X})
\label{eq:mutual2}
\end{eqnarray}
$N_{d_1}$ and $N_{d_2}$ are the the number of voxels for grid sizes
$d_1$ and $d_2$ respectively. The joint probability
$P(X_i,\bar{X}_k)=\frac{n_{ik}}{N}$, where $n_{ik}$ is the number of
galaxies shared by the $i^{th}$ voxel of grid size $d_1$ and $k^{th}$
voxel of grid size $d_2$. The joint entropy $H(X,\bar{X})$ is
similarly defined as,
\begin{eqnarray}
H(X,\bar{X}) & = & - \sum^{N_{d_1}}_{i=1} \sum^{N_{d_2}}_{k=1} \, P(X_i,\bar{X}_k) \, \log\, P(X_i,\bar{X}_k)
\label{eq:joint2}
\end{eqnarray}
We vary the grid size and estimate $I(X;\bar{X})$ for each choice of $(X,\bar{X})$.

\subsection{The conditional mutual information between morphology and environment}
A positive mutual information between morphology and environment does
not necessarily imply a causal dependence of morphology on
environment. The mutual information between morphology and environment
may come from the shared mutual information between different
environmental properties such as the environments on different
scales. To investigate this further, we calculate the conditional
mutual information between the morphology and environment.

The conditional mutual information provides the expected value of the
mutual information between two random variables given that we have the
knowledge of a third random variable. In particular here we are
interested in the mutual information between galaxy morphology and the
environment on a particular scale given that we have the knowledge of
the environment of the galaxy on a different scale. The conditional
mutual information between $X$ and $Y$ given the value of $\bar{X}$ is
defined as, \setlength\arraycolsep{1 pt}
\begin{eqnarray}
\nonumber I(X;Y|\bar{X})&=&\sum^{N_{d_1}}_{i=1}\sum^{2}_{j=1}\sum^{N_{d_2}}_{k=1}\,P(X_i,Y_j,\bar{X}_{k}) \, \log\,\frac{P(\bar{X}_k)P(X_i,Y_j,\bar{X}_{k})}{P(X_i,\bar{X}_{k})P(Y_j,\bar{X}_{k})}\\
&=&H(X,\bar{X})+H(Y,\bar{X})-H(X,Y,\bar{X})-H(\bar{X})
\label{eq:conditional}
\end{eqnarray}

Applying the chain rule of probability one can write the joint
probability
$P(X_i,Y_j,\bar{X}_{k})=P(\bar{X}_k|Y_j,X_i)P(Y_j|X_i)P(X_i)$ which
will be $\frac{(n_s)_{ik}}{N}$ for $j=1$ and $\frac{(n_e)_{ik}}{N}$
for $j=2$. $(n_s)_{ik}$ and $(n_e)_{ik}$ are respectively the numbers
of spirals and ellipticals shared by the $i^{th}$ voxel of size $d_1$
and $k^{th}$ voxel of size $d_2$. The joint entropy $H(X,Y,\bar{X})$
is defined as, \setlength\arraycolsep{1 pt}
\begin{eqnarray}
\nonumber H(X,Y,\bar{X}) & = & - \sum^{N_{d_1}}_{i=1} \sum^{2}_{j=1}\sum^{N_{d_2}}_{k=1} \, P(X_i,Y_j,\bar{X}_k) \, \log\, P(X_i,Y_j,\bar{X}_k)\\
&&
\label{eq:joint2}
\end{eqnarray}
The conditional mutual information is always non-negative. It is zero
only when the correlation between $X$ and $Y$ is entirely due to the
influence of $\bar{X}$. We vary the grid size and estimate
$I(X;Y|\bar{X})$ for each choice of $(X,\bar{X})$.

\subsection{The interaction information between morphology and environment}
One can define the interaction information \citep {mcgill} as,
\begin{eqnarray}
I(X;Y;\bar{X})&=&I(X;Y|\bar{X})-I(X;Y) 
\label{eq:interaction}
\end{eqnarray}
which measures the loss or gain in information between $X$ and $Y$ due
to the additional knowledge of the third random variable $\bar{X}$.
Conditioning on $\bar{X}$ may increase, decrease or not change the
mutual information between $X$ and $Y$. Consequently, $I(X;Y;\bar{X})$
may be positive, negative or zero. $I(X;Y;\bar{X})<0$ indicates that
conditioning on $\bar{X}$ weakens the correlation between $X$ and $Y$
whereas $I(X;Y;\bar{X})>0$ indicates that $\bar{X}$ enhances the
correlation between $X$ and $Y$. If conditioning on $\bar{X}$ does not
affect the mutual information between $X$ and $Y$ then
$I(X;Y;\bar{X})=0$.
 
We have used a $\Lambda$CDM cosmological model with
$\Omega_{m0}=0.31$, $\Omega_{\Lambda0}=0.69$ and $h=1$ throughout.

\begin{figure*}
\hspace*{-0.95cm}
\resizebox{5.5cm}{!}{\rotatebox{0}{\includegraphics{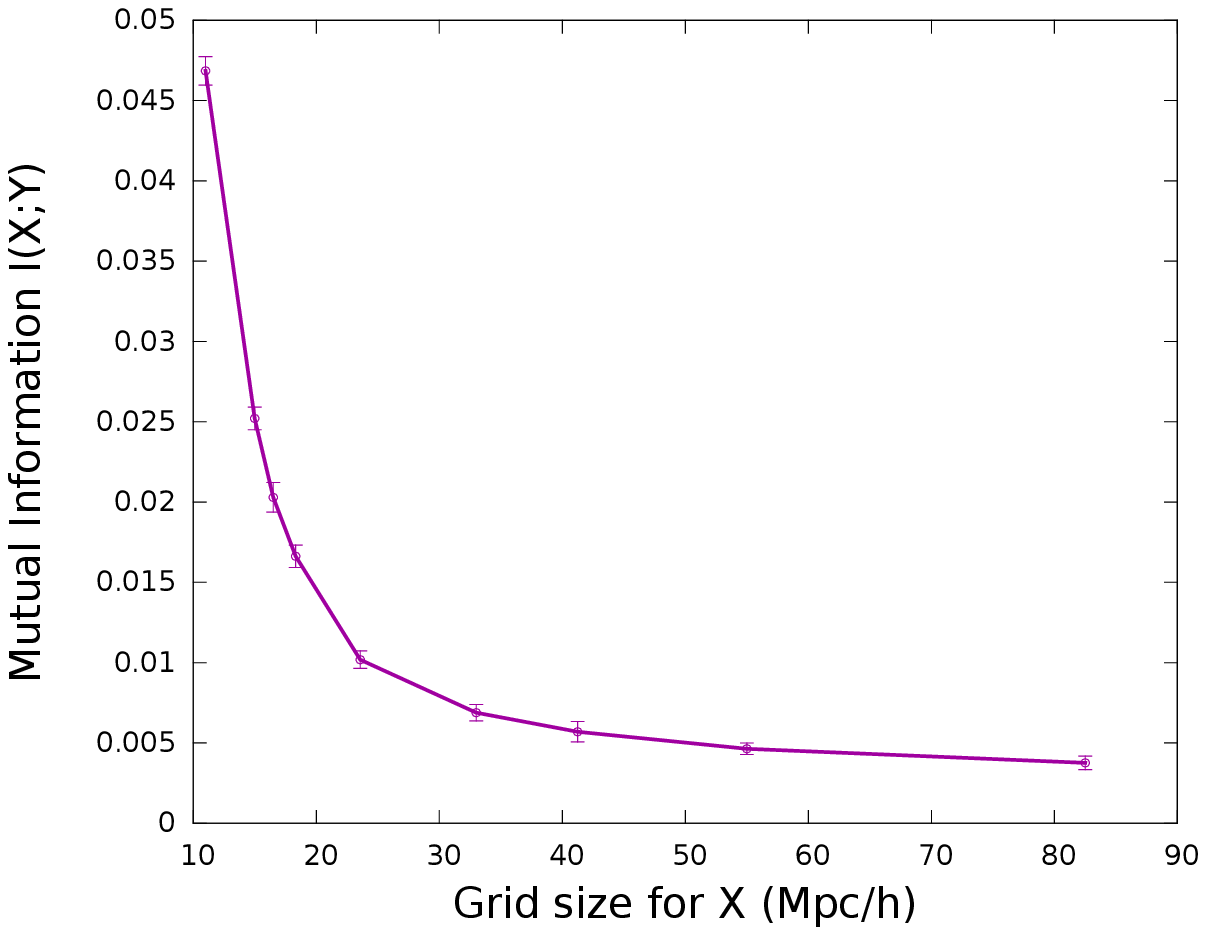}}}%
\hspace*{1cm}
\resizebox{5.5cm}{!}{\rotatebox{0}{\includegraphics{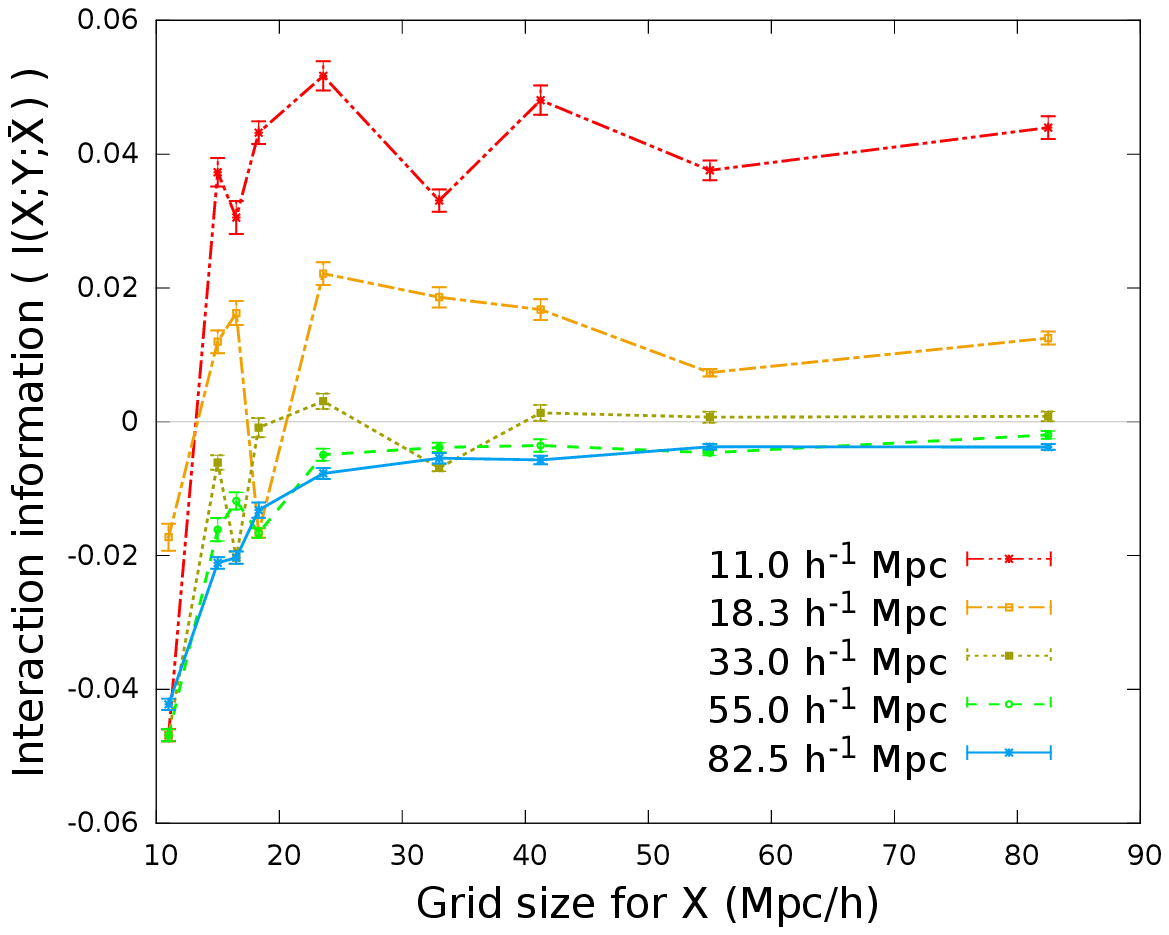}}}\\
\resizebox{6.7cm}{!}{\rotatebox{0}{\includegraphics{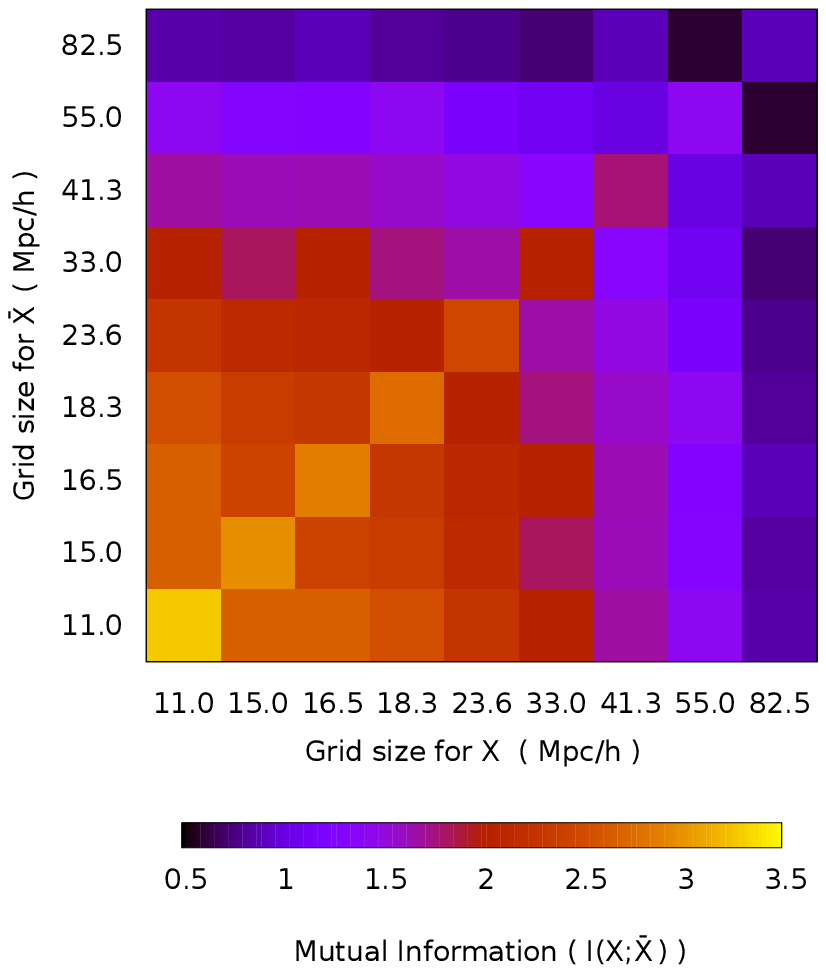}}}%
\resizebox{6.7cm}{!}{\rotatebox{0}{\includegraphics{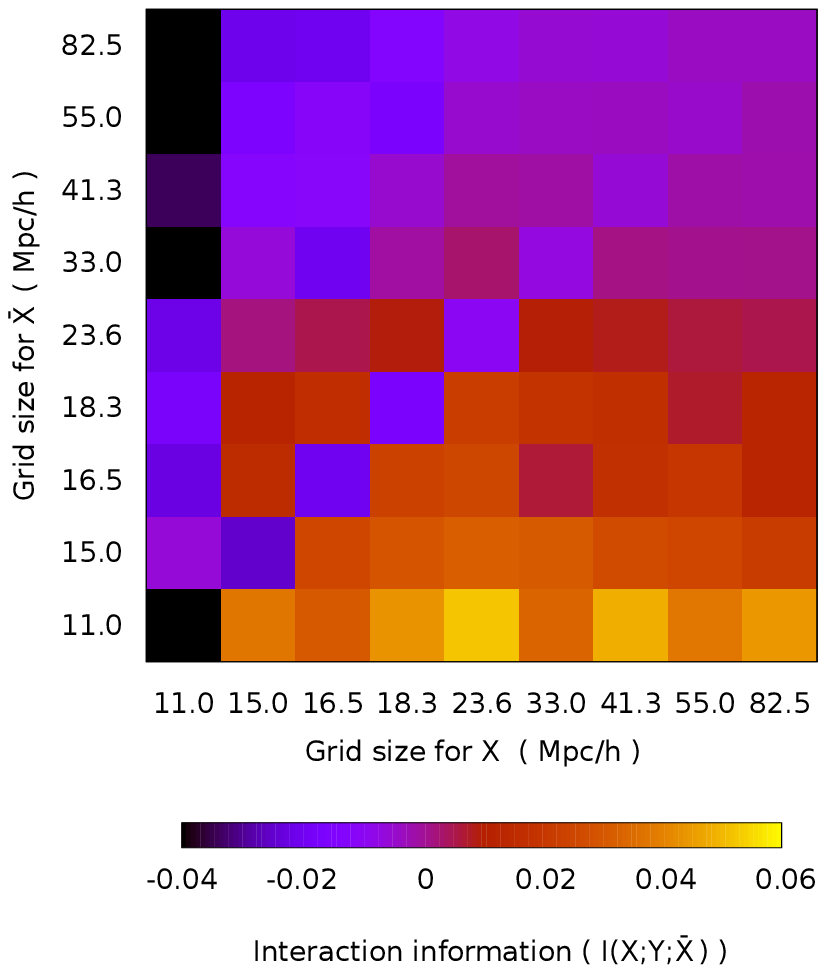}}}\\
\caption{ The top left panel shows the mutual information between
  morphology ($Y$) and environment ($X$) as a function of length
  scales. The top right panel shows the magnitudes of interaction
  information between morphology ($Y$) and environment ($X$) as a
  function of length scales when the environment is conditioned at
  different scales ($\bar{X}$) as indicated in the panel. The
  $1\sigma$ errorbars shown in the top two panels are obtained from
  $10$ jackknife subsamples drawn from the SDSS data. The interaction
  information is shown again for all possible combinations of
  $X,\bar{X}$ in the bottom right panel. The bottom left panel shows
  the mutual information between the environments on different scales
  ($X$ and $\bar{X}$). }
  \label{fig:information}
\end{figure*}

\section{DATA: THE GALAXY ZOO SAMPLE}
The Galaxy Zoo \citep{lintott1,lintott2} data can be directly accessed
from the SDSS DR12 SkyServer database. We download the data from the
SDSS DR12 database using a Structured Query Language (SQL) search. We
identify a contiguous region spanning $135^{\circ} \leq \alpha \leq
225^{\circ}$ and $0^{\circ} \leq \delta \leq 60^{\circ}$ where
$\alpha$ and $\delta$ are the right ascension and declination
respectively. We extract all the galaxies in this contiguous region
which lies in the redshift range $0 \leq z \leq 0.25$ and are brighter
than a limiting r-band Petrosian magnitude $17.77$. These cuts produce
a sample with $343340$ galaxies of which $134745$ are classified. We
only consider the classified galaxies with a debiased vote fraction
$>0.8$. We construct a volume limited sample by restricting the
extinction corrected and k-corrected r-band absolute magnitude to
$M_{r}<-20$. The resulting volume limited sample extends upto a
redshift $z=0.1067$ and consists of $42334$ galaxies with visual
morphological classification. It may be noted here that we have only
considered the galaxies which are visually classified as spiral or
elliptical and discarded all the galaxies with uncertain morphology.
Finally from this volume limited sample we extract a cubic region with
sides $165 \hmpc$ which contain a total $15860$ galaxies out of which
$11875$ are spirals and $3985$ are ellipticals.

\section{RESULTS AND CONCLUSIONS}
We show the mutual information $I(X;Y)$ between the morphology and
environment of galaxies as a function of length scales in the top left
panel of figure \autoref{fig:information}. It shows the presence of a
non-zero mutual information between morphology and environment which
decreases with increasing length scales but persists throughout the
entire length scales probed. The $1\sigma$ errorbars shown here are
obtained using $10$ jackknife subsamples drawn from the data. It is
important to understand the origin of this non-zero mutual information
between morphology and environment. To investigate this further, we
compute the conditional mutual information $I(X;Y|\bar{X})$ between
morphology and environment at each length scales by conditioning the
environment at other length scales. The mutual and the conditional
mutual information were then used to calculate the interaction
information $I(X;Y;\bar{X})$ shown in the bottom right panel of
\autoref{fig:information}. We find that conditioning the environment
on $11 \hmpc$ leads to a positive interaction between morphology and
environment at all the other scales. As we condition the environment
on successively larger scales we observe the same trend upto $33
\hmpc$. The interaction information becomes negative at all length
scales when the environment is conditioned on scales beyond $33
\hmpc$. It may be noted that all the diagonal boxes exhibit negative
interactions. Conditioning the environment on the same scale where the
interaction is measured would give $I(X;Y|\bar{X})=0$ leading to a
negative interaction information. In the top right panel of
\autoref{fig:information} we again separately show the magnitudes of
the interaction information at each length scale when the environment
is conditioned at different scales. This again clearly shows that the
degree of positive interaction between the morphology ($Y$) and
environment ($X$) decreases with an increase in the scale at which the
environment ($\bar{X}$) is conditioned. The interaction information
diminishes nearly to $0$ when the environment ($\bar{X}$) is
conditioned at $33 \hmpc$. The $1\sigma$ errorbars in each case are
obtained by jackknife resampling. The size of the errorbars indicate
that the interaction information are quite distinct when the
conditioning is done at a length scale below $33 \hmpc$. As noted
earlier, a further increase in the scale for conditioning $\bar{X}$
leads to a negative interaction between the morphology and environment
on all scales. It may be noted that the grid sizes for $X$ and
$\bar{X}$ are decided by the choice of the number of grids used to
divide the cubic region.

A positive interaction information imply `synergy' and a negative
interaction information imply `redundancy' in the interaction between
the random variables \citep{mcgill}. So a positive interaction between
$X$ and $Y$ given the knowledge of $\bar{X}$ indicates that when
$\bar{X}$ is known, the knowledge of $X$ provides additional
information on $Y$ than that provided by $X$ and $\bar{X}$
individually. On the other hand, a negative interaction between $X$
and $Y$ given the knowledge of $\bar{X}$ tells us that when $\bar{X}$
is known, the knowledge of $X$ does not tell us anything new about
$Y$. Conditioning $\bar{X}$ upto a length scales of $33 \hmpc$
introduces positive interaction between $X$ and $Y$ on larger scales
and negative interaction on smaller scales. This indicates that
conditioning the environment on a particular scale plays an important
role in the apparent interaction between morphology and environment.
The fact that the interaction between $X$ and $Y$ becomes negative on
all scales when $\bar{X}$ is conditioned on a scale beyond $33 \hmpc$
suggests that no information is shared between morphology and
environment beyond this length scale. Further the mutual information
between $X$ and $Y$ on scales below $33 \hmpc$ does not arise due to a
causal interaction between them. Rather they are the outcome of a
non-zero mutual information between the environments on different
scales. \autoref{eq:mutual2} and \autoref{eq:conditional} together
suggest that a larger mutual information between the environments on
different scales would produce a smaller conditional mutual
information between $X$ and $Y$ given $\bar{X}$. Consequently the
interaction information $I(X;Y;\bar{X})$ would reduce when both
$I(X;\bar{X})$ and $I(X;Y)$ are larger and enhance when $I(X;\bar{X})$
is larger but $I(X;Y)$ is smaller. In the bottom left panel of
\autoref{fig:information} we show the mutual information
$I(X,\bar{X})$ shared by the environments on different scales. It is
interesting to note that the mutual information between $X$ and
$\bar{X}$ are much larger than the mutual information between $X$ and
$Y$ on each scale. Furthermore the mutual information between $X$ and
$\bar{X}$ decreases with increasing length scales and reduces by more
than a factor of half beyond a scale of $33 \hmpc$. Combining these
results we conclude that the observed apparent interaction between
morphology and environment of a galaxy are most likely caused by the
shared mutual information between the environments on different scales
and they cease to exist beyond a length scales of $\sim 30 \hmpc$.

Finally we note that the method presented here provides an effective
avenue to explore the large scale environmental dependence of galaxy
properties from a new perspective. In future, we plan to carry out a
detail study of the large scale environmental dependence of a number
of other galaxy properties using the SDSS \citep{york}. Such a study
will clearly reveal any influence of the large-scale environment on
the other galaxy properties and their evolution.

\section{ACKNOWLEDGEMENT}
The authors would like to thank the Galaxy Zoo team for making the
data public. The authors thank an anonymous referee for useful
comments on the paper. B.P. would like to acknowledge financial
support from the SERB, DST, Government of India through the project
EMR/2015/001037. B.P. would also like to acknowledge IUCAA, Pune and
CTS, IIT, Kharagpur for providing support through associateship and
visitors programme respectively. S.S. thanks Chris J. Lintott for his
help in understanding the Galaxy Zoo data. S.S. would also like to
thank UGC, Government of India for providing financial support through
a Rajiv Gandhi National Fellowship.

\bsp	
\label{lastpage}
\end{document}